\documentclass[showpacs,preprintnumbers,amsmath,amssymb,superscriptaddress]{revtex4}
\usepackage{graphicx}% Include figure files
\usepackage{dcolumn}% Align table columns on decimal point
\usepackage{bm}% bold math
\newcommand{\UA}{\mbox{$U_{A}(1)$}}

\newcommand{\ie}{{\it i.e.}}
\newcommand{\bea}{\begin{eqnarray}}
\newcommand{\eea}{\end{eqnarray}}
\def\JLone<#1,#2>{#1}
\def\JLtwo<#1,#2,#3>{#2}
\def\JLyear<#1,#2,#3,#4>{#3}
\def\JLpage<#1,#2,#3,#4>{#4}
\newcommand\JL[1]{\JLone<#1>\ {\bfseries \JLtwo<#1>}, \JLpage<#1> (\JLyear<#1>)}
\def\Jpage<#1,#2,#3>{#3}

\newcommand\andvol[1]{{\bfseries \JLone<#1>}, \Jpage<#1> (\JLtwo<#1>)}

\newcommand\PTP[1]{Prog.\ Theor.\ Phys.\ \andvol{#1}}

\newcommand\PRC[1]{Phys.\ Rev.\ C\ \andvol{#1}}
\newcommand\PRD[1]{Phys.\ Rev.\ D\ \andvol{#1}}

\newcommand\PRL[1]{Phys.\ Rev.\ Lett.\ \andvol{#1}}
\newcommand\PLB[1]{Phys.\ Lett.\ B\ \andvol{#1}}
\newcommand\NPA[1]{Nucl.\ Phys.\ A\ \andvol{#1}}
\newcommand\NPB[1]{Nucl.\ Phys.\ B\ \andvol{#1}}
\newcommand\JP[1]{J.~of Phys.\ \andvol{#1}}
\newcommand\ANN[1]{Ann.\ of Phys.\ \andvol{#1}}

\newcommand\PRP[1]{Phys. Rep.\ \andvol{#1}}

\def\citen#1{%
\if@filesw \immediate \write \@auxout {\string \citation {#1}}\fi 
\@tempcntb\m@ne \let\@h@ld\relax \def\@citea{}%
\@for \@citeb:=#1\do {%
  \@ifundefined {b@\@citeb}%
    {\@h@ld\@citea\@tempcntb\m@ne{\bf ?}%
    \@warning {Citation `\@citeb ' on page \thepage \space undefined}}%
    {\@tempcnta\@tempcntb \advance\@tempcnta\@ne
    \setbox\z@\hbox\bgroup % check if citation is a number:
    \ifnum0<0\csname b@\@citeb \endcsname) \relax
       \egroup \@tempcntb\number\csname b@\@citeb \endcsname \relax
       \else \egroup \@tempcntb\m@ne \fi
    \ifnum\@tempcnta=\@tempcntb % Number follows previous--hold on to it
       \ifx\@h@ld\relax % first pair of successives
          \edef \@h@ld{\@citea\csname b@\@citeb\endcsname)}% 
       \else % compressible list of successives
          \edef\@h@ld{\penalty\@highpenalty\hskip.15em plus.1em minus.1em
            \hbox{-}\penalty\@highpenalty\hskip.15em plus.1em minus.1em
            \csname b@\@citeb\endcsname)}%
       \fi
    \else   %  non-successor--dump what's held and do this one
       \@h@ld\@citea\csname b@\@citeb \endcsname)\let\@h@ld\relax
    \fi}%
 \def\@citea{,\penalty\@highpenalty\hskip.15em plus.1em minus.1em}%
}\@h@ld}
\def\@citex[#1]#2{\@cite{\citen{#2}}{#1}}%
\def\@cite#1#2{\leavevmode\unskip
  \ifnum\lastpenalty=\z@\penalty\@highpenalty\fi% highpenalty before
   $^{\hskip.15em plus.1em minus.1em \multiply\@highpenalty 3 #1%
      \if@tempswa,\penalty\@highpenalty\ #2\fi % and before note.
    }$\spacefactor\@m}
\begin{document}
%\preprint{APS/123-QED}

\title{Light Scalar Mesons in the Improved Ladder QCD}% Force line breaks with \\

\author{Toru Umekawa}
 \affiliation{
 Department of Physics, Tokyo Institute of Technology, Meguro,\\
 Tokyo 152-8551 Japan\\}

\author{Kenichi Naito}%
\affiliation{%
Meme Media Laboratory, Graduate school of Enginieering, 
Hokkaido University,\\
Sapporo, Hokkaido 060-8628 Japan\\
}%

\author{Makoto Oka}
 \affiliation{
 Department of Physics, Tokyo Institute of Technology, Meguro,\\
 Tokyo 152-8551 Japan\\}

\author{Makoto Takizawa}
\affiliation{
Showa Pharmaceutical University, Machida, Tokyo 194-8543 Japan
}%

\date{\today}% It is always \today, today,

\begin{abstract}
The light scalar meson spectrum is studied using the improved ladder QCD 
with the \UA\ breaking Kobayashi-Maskawa-'t~Hooft interaction by 
solving the Schwinger-Dyson and Bethe-Salpeter equations. 
The dynamically generated momentum-dependent quark mass is large enough 
in the low momentum region to give rise to the spontaneous breaking of 
chiral symmetry. 
Due to the large dynamical quark mass, the scalar mesons become the
$q\bar{q}$ bound states. 
Since the parameters have been all fixed to reproduce the light 
pseudoscalar meson masses and the decay constant, 
there is no free parameter in the calculation of the scalar mesons.
We obtain $M_{\sigma} = 667$ MeV, $M_{a_0} = 942$ MeV and 
$M_{f_0} = 1336$ MeV. They are in good agreement with the observed masses
of $\sigma(600)$, $a_0(980)$ and $f_0(1370)$, respectively.
We therefore conclude that these states are the members of the 
light scalar meson nonet.
The mass of $K_0^{*}$ is obtained between that of $a_0$ and $f_0$ and
the corresponding state is not observed experimentally.  
We also find that the strangeness content in the $\sigma$ meson 
is about 5\%.
\end{abstract}

\maketitle

%\section{\label{sec:level1}First-level heading:\protect\\ The line
%break was forced \lowercase{via} \textbackslash\textbackslash}
%This sample document demonstrates proper use of REV\TeX~4 (and
%\LaTeXe) in mansucripts prepared for submission to APS
%journals. Further information can be found in the REV\TeX~4
%documentation included in the distribution or available at
%\url{http://publish.aps.org/revtex4/}.
%
%When commands are referred to in this example file, they are always
%shown with their required arguments, using normal \TeX{} format. In
%this format, \verb+#1+, \verb+#2+, etc. stand for required
%author-supplied arguments to commands. For example, in
%\verb+\section{#1}+ the \verb+#1+ stands for the title text of the
%author's section heading, and in \verb+\title{#1}+ the \verb+#1+
%stands for the title text of the paper.
%
%Line breaks in section headings at all levels can be introduced using
%\textbackslash\textbackslash. A blank input line tells \TeX\ that the
%paragraph has ended. Note that top-level section headings are
%automatically uppercased. If a specific letter or word should appear in
%lowercase instead, you must escape it using \verb+\lowercase{#1}+ as
%in the word ``via'' above.

%%%%%%%%%%%%%%%%%%%%%%%%%%%%%%%%%%%%%%%%%%%%%%%%%%%%%%%%%%%%%%%%%%%
\section{\label{sec:intro}Introduction}
%%%%%%%%%%%%%%%%%%%%%%%%%%%%%%%%%%%%%%%%%%%%%%%%%%%%%%%%%%%%%%%%%%%

About three decades ago, quantum chromodynamics (QCD) was conceived as 
the microscopic theory of the strong interactions. 
Since then, many aspects of QCD have been studied and QCD is established 
as the fundamental theory of the strong interactions.
Because of the nonperturbative nature of the low-energy QCD, 
understanding the low-lying hadron structures from the viewpoint of 
the quark and gluon degree of freedom is one of the most challenging problems.
\par
The lightest excitation on the QCD vacuum is the pion which is considered as 
a quark and an antiquark bound state in the pseudoscalar channel. 
Its mass (about 140 MeV) is off-scale light compared with other hadrons 
such as the $\rho$-meson ($m_\rho \sim 770$ MeV), the spin-flip partner 
of the pion, and the nucleon ($m_N \sim 940$ MeV), the three-quark bound state.
This can be understood by recognizing that the chiral symmetry is 
spontaneously broken in the QCD vacuum and the pion is the Nambu-Goldstone (NG)
boson associated with the dynamical chiral symmetry breaking (DCSB).
The DCSB is, therefore, among the most important aspects of low-energy hadron
physics.  It is believed to be responsible for a large part of the constituent 
quark masses, which are introduced in the many constituent quark models.
\par
If QCD lagrangian has no explicit chiral-symmetry breaking term, the mass of 
the NG boson associated with the DCSB should be zero.  In order to explain 
the observed mass of the pion, one needs small explicit chiral-symmetry 
breaking terms, namely, current quark mass terms.
The NG boson nature of the pion was first studied by Nambu and Jona-Lasinio 
(NJL) using the schematic model with the four-fermion interaction \cite{NJL}.
Since then, low-energy hadron properties have been widely studied using 
the NJL-type models \cite{HK1994}.
On the other hand, the effects of the explicit breaking of the chiral symmetry
on the pion properties have been systematically studied using the effective 
lagrangian composed of the pion field. This approach is called the chiral
perturbation theory (ChPT) \cite{GL1984}. The success of ChPT approach 
supports the importance of the DCSB in low-energy QCD.
\par
When we start looking at the strange quark sector, we meet another problem.
If one assumes that the up, down and strange quark masses are small compared 
with the scale of the DCSB, the number of the NG bosons is equal to the 
dimension of the coset space $U_L(3) \times U_R(3) / U_V(3)$, namely, nine.
The ninth heavier pseudoscalar meson is $\eta'$ and its mass is 
much heavier than the other octet pseudoscalar mesons. 
Weinberg showed that the mass of $\eta'$ should be less
than $\sqrt{3} m_{\pi}$ if \UA\ symmetry were not
explicitly broken \cite{Weinberg1975}.
Thus the \UA\ symmetry must be broken.
The key step to solve this \UA\ problem was to realize that there is an
anomaly in the \UA\ channel. Namely the \UA\ symmetry 
in the classical theory, i.e., in the action, is broken by quantum effects.
In the following year, 't Hooft pointed out the relation between 
\UA\ anomaly and topological gluon configurations of QCD and 
showed that the interaction of light quarks and instantons breaks 
the \UA\ symmetry\cite{tHooft1976}.
He also showed that such an interaction can be represented by a
local $2N_{f}$ quark vertex, which is antisymmetric under
flavor exchanges, in the dilute instanton gas approximation.
\par
The effects of the \UA\ anomaly on the low-energy QCD have been
extensively studied in the $1/N_C$ expansion approach \cite{Nc}.
In the $N_C \rightarrow \infty$ limit, the \UA\ anomaly is turned off
and then the $\eta$ and $\eta'$ mesons become the ideal mixing states.
The flavor component of the $\eta$ is $\frac{1}{2} (u \bar u + d \bar d)$ with 
$m_\eta(N_C \rightarrow \infty) = m_\pi$ and
the $\eta'$ meson becomes a pure $s \bar s$ state with 
$m_{\eta'}^2 (N_C \rightarrow \infty) \simeq 2 m_K^2 - m_\pi^2 = 
(687 \rm{MeV})^2$ \cite{Veneziano}. 
The higher oder effects of the $1/N_C$ expansion
give rise to the flavor mixing between the $\eta$ and $\eta'$ mesons 
and push up the $\eta'$ mass. They were further discussed in the 
context of the ChPT \cite{NcChPT} and the reasonable description of 
the nonet pseudoscalar mesons was obtained.
\par
The \UA\ breaking $2 N_f$ quark determinant interaction was 
introduced to the low-energy effective quark models of QCD.
The low-lying meson properties have been studied 
\cite{KH1988,BJM1988, KKT1988, RA1988,KLVW1990} using 
the three-flavor version of the NJL model with the Kobayashi-Maskawa-'t~Hooft 
(KMT) determinant interaction \cite{tHooft1976,KM1970}.  
The radiative decays of the $\eta$ meson have been 
studied in this approach \cite{TO1995} and found that 
these decay widths are reproduced when the \UA-breaking 
interaction is much stronger than the previous studies 
\cite{KH1988,BJM1988, KKT1988, RA1988,KLVW1990}.
It is further argued that the \UA-breaking interaction gives
rise to the spin-spin forces, which are important for light baryons 
\cite{SR1989,OT1989,MT1993}.
\par
The dynamics of instantons in the multi-instanton vacuum has
been studied by many authors, either in the models or in the
lattice QCD approach, and the widely accepted picture is that
the QCD vacuum consists of small instantons of the size about
1/3 fm with the density of 1 instanton (or anti-instanton) per
fm$^{4}$ \cite{Instanton}. In the instanton liquid model, 
the instanton plays a crucial role in understanding not only the 
$\UA$ anomaly but also the spontaneous breaking of chiral symmetry itself.
\par
Recently, the low-lying scalar mesons, $J^{\pi}=0^{+}$, attracts a lot of 
attention by two reasons.\cite{sigmaworkshop}
(1) Experimental evidence for $\sigma$ 
($I=0$) scalar meson of mass around 600 MeV is overwhelming 
\cite{PDG2002,Ishida1996,IH1999,KLL1999}.
Especially the decays of heavy mesons show clear peaks in the 
$\pi\pi$ invariant mass spectrum. Including this a rather light 
isoscalar state, the light scalar mesons show strange mass patterns, i.e., 
$\sigma(600) - a_0(980) - K_0^*(?) - f_0(980)$, where $K_0^*$ is not confirmed. This pattern cannot be explained as a naive $q\bar q$ nonet, 
because the $I = 1$ $a_0$ states are almost degenerate with the second 
$I = 0$ state $f_0$, while the first $I = 0$ state $\sigma$ is far below them.
(2) The roles of the scalar mesons in chiral symmetry have been 
stressed in the context of high temperature and/or density hadronic 
matter \cite{HK1985}.
It is believed that chiral symmetry is restored in the QCD ground 
state at high temperature (and/or baryon density).  Above the critical 
temperature, $\simeq 170$ MeV, the world is nearly chiral symmetric and 
we expect that hadrons belong to irreducible representations of 
chiral symmetry, if we neglect small mixing due to finite quark mass.
The pion is not any more a Nambu-Goldstone boson, and has a finite 
mass and should be degenerate with a scalar meson, \ie, sigma.
Another scenario was proposed recently. It is the vector manifestation 
where the rho meson becomes massless degenerate with pion as the chiral partner
\cite{HS2002}.  In order to make the situation clear, we consider that
the light scalar mesons are the key particles.
\par
We have studied the effects of the \UA\ breaking interaction on the low-lying 
nonet scalar mesons using the extended NJL model, in which the three-flavor 
NJL model is supplemented with the KMT determinant interaction \cite{NOTU2003}.
Why is the \UA\ expected to be important in the scalar mesons?  
It is because the KMT interaction selects out the scalar sector as 
well as the pseudoscalar mesons and therefore the OZI rule may be 
broken significantly also in the scalar mesons \cite{Dm1996}.
We have found that the \UA\ breaking interaction gives rise to about 150 MeV
mass difference between the $\sigma$ and $a_0$ mesons. 
We have also found that the strangeness content in the $\sigma$ meson is 
about 15\%. The calculated mass of the $I = 1/2$ state ($K_0^*$) was about 200
MeV heavier than that of the $I = 1$ state ($a_0$).
\par
The physics of the light scalar mesons seems to be directly related to 
the mechanism of the dynamical chiral symmetry breaking (DCSB). In the NJL 
model, the strong four-quark interaction gives rise to the DCSB and 
it leads the simple mass relation between the scalar meson mass ($m_S$) 
and the dynamical quark mass ($M_q$) in the mean-field approximation, i.e., 
$m_S = 2 M_q$ in the chiral limit.  In the case of the instanton liquid model,
the DCSB is caused by the instanton induced interaction. The scalar meson 
masses are rather sensitive to the instantion-antiinstanton interaction.
The results in the fully interacting instanton ensemble are 
$m_\sigma \simeq 0.58$ GeV and $m_{a_0} \simeq 2.05$ GeV \cite{Instanton}.
\par
In contrast with the instanton liquid model, the study of the 
QCD Schwinger-Dyson (SD) equation for the quark propagator in the 
improved ladder approximation (ILA) has shown that the spontaneous
breaking of the chiral symmetry is explained by simply extrapolating 
the running coupling constant from the perturbative high-energy region to 
the low-energy region \cite{Higashijima}. 
Then, the Bethe-Salpeter (BS) equation for the $J^{PC} = 0^{-+}$ 
$q \bar q$ channel has been solved in the ILA and the existence of 
the Nambu-Goldstone pion has been confirmed \cite{ILA,NYNOT1999-0}.
The numerical predictions of the pion decay constant $f_{\pi}$ and the
quark condensate $\langle \overline{\psi} \psi \rangle$ are rather good.
It has been also shown that the BS amplitude shows the correct
asymptotic behavior as predicted by the operator product expansion (OPE) 
in QCD \cite{ILA1}.  
The masses and decay constants for the lowest lying scalar, vector and 
axial-vector mesons have been evaluated by calculating
the two point correlation functions for the composite operators
$\overline{\psi} M \psi$.  The obtained values are in reasonable agreement with
the observed ones \cite{ILA2}.  
\par
Then, the current quark mass term has 
been introduced in the studies of the BS amplitudes in the ILA \cite{NYNOT1999}
and the reasonable values of the pion mass, the pion decay constant and 
the quark condensate have been obtained. 
It has been also shown that the pion mass square and the pion decay 
constant are almost proportional to the current quark mass up to 
the strange quark mass region.
The effect of the \UA\ anomaly is further introduced in the ILA approach by the
KMT interaction.  The instanton size effects are taken into account by the 
form factor of the interaction vertices. 
It guarantees the right asymptotic behavior of the solutions of the 
SD and BS equations. The properties of the $\eta$ and $\eta'$ mesons have been 
studied by solving the coupled channel BS equations and the reasonable values 
of $m_\pi$, $m_\eta$, $m_{\eta'}$, $f_\pi$ and 
$\langle \bar qq \rangle_R$ with a relatively weak
flavor mixing interaction (KMT interaction), 
for which the chiral symmetry breaking is dominantly induced by the 
soft-gluon exchange interaction \cite{NNTYO2000}.
\par
The purpose of this paper is to study the properties of the light scalar
meson nonet in the improved ladder approximation (ILA) of QCD with 
the \UA\ breaking Kobayashi-Maskawa-'t~Hooft (KMT) 6-quark flavor determinant
interaction.
In this approach, the mechanism of the dynamical chiral symmetry breaking is
different from the NJL model and instanton liquid model. 
It has been shown that the Wilsonian non-perturbative renormalization group 
equation gives the identical effective fermion mass with that obtained by 
solving the Schwinger-Dyson equation in the improved ladder approximation 
\cite{AMSTT2000}. We hope that the present study may shed light on the 
mechanism of the DCSB in the low-energy QCD.
It should be noted here that the parameters have been all fixed in the 
pseudoscalar meson sector and therefore there is no free parameter in the 
present study.
\par
There have been many studies of the pion BS amplitude using the 
effective models of QCD and /or the approximation schemes of QCD 
\cite{Roberts,Roberts2}.  The main differences among these studies are the 
form of the gluon propagator used in the SD and BS equations. 
Of course, the behavior of the gluon propagator in the asymptotic region
is well established and there is no differences. However, that in the
infrared region is not well known.  The simple infrared cutoff is introduced
in the ILA, while in the approach given in \cite{Roberts2}, 
the specific form of the infrared gluon propagator is assumed.
\par
Recently, attempts have been made to evaluate enhancement of the gluon 
propagator (as well as the vertex factor) from quenched lattice QCD data of 
the quark propagator \cite{IOS2003,Bha2003}.  They found that the gluon 
propagator is required to have strong enhancement in the soft momentum region 
so that chiral symmetry is dynamically broken.  Although the enhancement 
patterns vary among the models and parametrizations, they tend to show 
similar behavior when the solution of the SD equation, i.e., 
the quark mass function, is concerned. 
Thus we expect that the results 
for the mesonic 
bound state in the BS equation with the effective quark mass function 
qualitatively agree among the models. 
\par

The paper is organized as follows.  In Sec.~\ref{sec:formulation} we 
explain the Lagrangian we  use in the present study and derive
the SD equation for the quark propagator and the BS equation for the
scalar meson. Sec.~\ref{sec:result} is devoted to the numerical results.
Finally, summary and concluding remarks are given in Sec.~\ref{sec:conclusion}.
%
%%%%%%%%%%%%%%%%%%%%%%%%%%%%%%%%%%%%%%%%%%%%%%%%%%%%%%%%%%%%%%%%%%%
\section{\label{sec:formulation}Formulation}
%%%%%%%%%%%%%%%%%%%%%%%%%%%%%%%%%%%%%%%%%%%%%%%%%%%%%%%%%%%%%%%%%%%
%
In this section, we present the formulation of 
the improved  ladder QCD with KMT interaction. 
The rainbow approximation is applied to the SD 
equation for the quark propagator and to the 
BS equation for the pseudoscalar and scalar mesons. 
Since the derivations of the SD equation and the BS equation 
for the pseudoscalar mesons have been given in the ref. \cite{NNTYO2000}, 
here, we present only the results. 
%
%%%%%%%%%%%%%%%%%%%%%%%%%%%%%%%%%%%%%%%%%%%%%%%%%%%%%%%%%%%%%%%%%%%
\subsection{\label{sec:ILA}Improved ladder QCD with KMT interaction}
%%%%%%%%%%%%%%%%%%%%%%%%%%%%%%%%%%%%%%%%%%%%%%%%%%%%%%%%%%%%%%%%%%%%
%
The improved ladder QCD is based on the ladder approximation which is 
improved by replacing the coupling constant by the running coupling constant. 
We employ the rainbow approximation of the SD equation for the quark 
propagator and the ladder approximation of the BS equation for the 
quark-antiquark bound states. 
Improvement is made by the use of the running coupling constant 
according to the Higashijima-Miransky \cite{Higashijima,ILA,ILA2} method. 
\par
Under this approximation, the gluon exchange interaction 
${\mathcal L}_{\rm GE}$ becomes
\begin{eqnarray}
{\mathcal L} _{\rm GE}
&=&
-\frac{1}{2} 
\int_{pp'qq'} 
i\bar{g}^2 \left( (\frac{p - q'}{2})^2, (\frac{q - p'}{2})^2 \right)
D^{\mu \nu}\left(\frac{p+p'}{2}-\frac{q+q'}{2} \right)
\nonumber \\
&&\qquad \qquad\times
\bar{\psi}(p)\gamma_{\mu}T^a\psi(p')
\bar{\psi}(q)\gamma_{\nu}T^a\psi(q')
e^{-i(p+p'+q+q')x}
.
\nonumber\\
\end{eqnarray}
where we use an abbreviation 
$\int_p = \int \frac{d^4p}{(2\pi)^4}$. 
For the gluon propagator we employ the Landau gauge, 
\begin{eqnarray}
iD^{\mu \nu}(k) = \left( g^{\mu \nu} -\frac{k^{\mu}k^{\nu}}{k^2} \right)
\frac{-1}{k^2}
\end{eqnarray}
and the Higashijima-Miransky type running coupling constant $\bar{g}^2$ 
defined as follows \cite{Higashijima,ILA,ILA2}.
\begin{eqnarray}
\bar{g}^2(p^2_E,q^2_E) 
= \theta(p^2_E-q^2_E)g^2(p^2_E)+ \theta(q^2_E-p^2_E)g^2(q^2_E)
\end{eqnarray}
with
\begin{eqnarray}
\label{runningCoupling}
&&g^2(p_E^2) = 
\left\{
\begin{array}{ll}
\frac{1}{\beta_0}\frac{1}{1+t}  
&
\mbox{for} \quad t_{\mathrm{IF}} \le t \\
&\\
\frac{1}{2\beta_0}\frac{3 t_{\mathrm{IF}} -t_0 +2 
      -\frac{(t-t_0)^2}{t_{\mathrm{IF}}-t_0}
}{(1+t_{\mathrm{IF}})^2}
 & \mbox{for} \quad t_0 \le t \le t_{\mathrm{IF}} \\
&\\
\frac{1}{2\beta_0}
\frac{3 t_{\mathrm{IF}} -t_0 +2}{(1+t_{\mathrm{IF}})^2}
 & \mbox{for} \quad t \le t_0  \\
\end{array}
\right. 
\\
&&t=\ln \frac{p_E^2}{\Lambda_{\mathrm{QCD}}^2} -1 \\
&&\beta_{0} = \frac{1}{(4 \pi)^2}\frac{11N_C-2N_f}{3}
\end{eqnarray}
Here, $p_E$ and $q_E$ denote the Euclidian momenta defined by 
\begin{eqnarray}
p = (p_0, \vec{p}) \quad &\to& \quad p_E = (ip_4, \vec{p}) \\
p^2 = p_0^2 - \vec{p}^2 \quad &\to& \quad p^2_E = -p^2 = \vec{p}^2 + p_0^2 
\end{eqnarray}
In eq.(\ref{runningCoupling}) the infrared cut-off $t_{\mathrm{IF}}$ 
is introduced. Above  $t_{\mathrm{IF}}$, $g^2(p_E^2)$ develops according to 
the one loop solution of the QCD  renormalization group equation, 
while below $t_0$, $g^2(p_E^2)$ is kept constant. 
These two regions are connected by a quadratic polynomial so that 
$g^2(p_E^2)$ becomes a smooth function.
Here $N_C$ and $N_f$ are the number of colors and active flavors respectively. 
We use $N_C=N_f=3$ in this study. 
\par
The KMT interaction ${\mathcal L} _{\rm KMT}$ is given by 
\begin{eqnarray}
{\mathcal L}_{\rm KMT}
&=&-\frac{1}{3}G_D\epsilon^{f_1f_2f_3}\epsilon^{g_1g_2g_3}
%\nonumber \\
%&\times&
\int_{p_1p_2p_3q_1q_2q_3}e^{-i(p_1+p_2+p_3+q_1+q_2+q_3)x} 
%\nonumber \\
%&\times&
w(p_1,p_2,p_3,q_1,q_2,q_3)
\nonumber \\
&\times&
\left\{
[\bar{\psi}_{g_1}(p_1)\psi_{f_1}(q_1)][\bar{\psi}_{g_2}(p_2)\psi_{f_2}(q_2)]
[\bar{\psi}_{g_3}(p_3)\psi_{f_3}(q_3)]
\right.
\nonumber \\
&&+3
\left.
[\bar{\psi}_{g_1}(p_1)\psi_{f_1}(q_1)][\bar{\psi}_{g_2}(p_2)\gamma_5\psi_{f_2}(q_2)]
[\bar{\psi}_{g_3}(p_3)\gamma_5\psi_{f_3}(q_3)]
\right\}
%\nonumber \\
\end{eqnarray}
where $f_1,g_1,\cdots$ are flavor indices, $\epsilon $ denotes 
the antisymmetric tensor with $\epsilon^{uds}=1$ 
This interaction breaks the $U_A(1)$ symmetry and also 
mixings quark flavors. 
We introduce a weight function 
\begin{eqnarray}
w(p_1,\cdots,q_3)=\exp(-\kappa(p_1^2 +\cdots + q_3^2))
\end{eqnarray}
so that the KMT interaction is turned off at large momentum region. 
Then the asymptotic behavior of the ILA are kept
consistent with the perturbative QCD. 
The parameter $\kappa$ is taken as $\kappa  = 0.7 [\mbox{GeV}^{-2}] $. 
This value corresponds to the form factor of the instanton 
of the average size $\rho \sim 1/3 \,[{\rm fm}]$ 
\cite{NNTYO2000}. 
%
%%%%%%%%%%%%%%%%%%%%%%%%%%%%%%%%%%%%%%%%%%%%%%%%%%%%%%%%%%%%%%%%%%%
\subsection{\label{sec:SD}Schwinger-Dyson Equation}
%%%%%%%%%%%%%%%%%%%%%%%%%%%%%%%%%%%%%%%%%%%%%%%%%%%%%%%%%%%%%%%%%%%%
%
The SD equation in the momentum space is written as 
\begin{figure*}
\includegraphics[height=2cm]{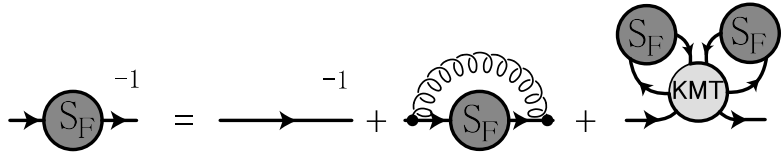}% Here is how to import EPS art
\caption{\label{fig:SDeq} The Schwinger-Dyson equation}
\end{figure*}
\begin{eqnarray}
%&&
i{S_{F}}^{-1}(q) - i {S_{0}}^{-1} (q)
%\nonumber \\
&=&
- C_F\int_{p}
\bar{g}^2(-q^2,-p^2)iD^{\mu\nu}(p-q)
\gamma_{\mu}{S_F}(p)\gamma_{\nu}
\nonumber\\
&&-G_D
\int_{p,k} w(-q^2-p^2-k^2)
%\nonumber\\
%&&\qquad\qquad\qquad \times
  \mbox{tr}^{\rm (DC)}[{S_F(p)}]
         \mbox{tr}^{\rm (DC)}[{S_F(k)}].
\nonumber \\
\end{eqnarray}
where tr$^{\rm (DC)}$ means the trace of the Dirac and color components 
and $C_F = (N_C^2 -1) /2N_C$. Here, the flavor antisymmetrization is assumed 
in the second term of the RHS.
This equation is shown diagrammatically in Fig. \ref{fig:SDeq}. 
\par
\begin{figure*}
\begin{tabular}{ccc}
\includegraphics[height=5cm]{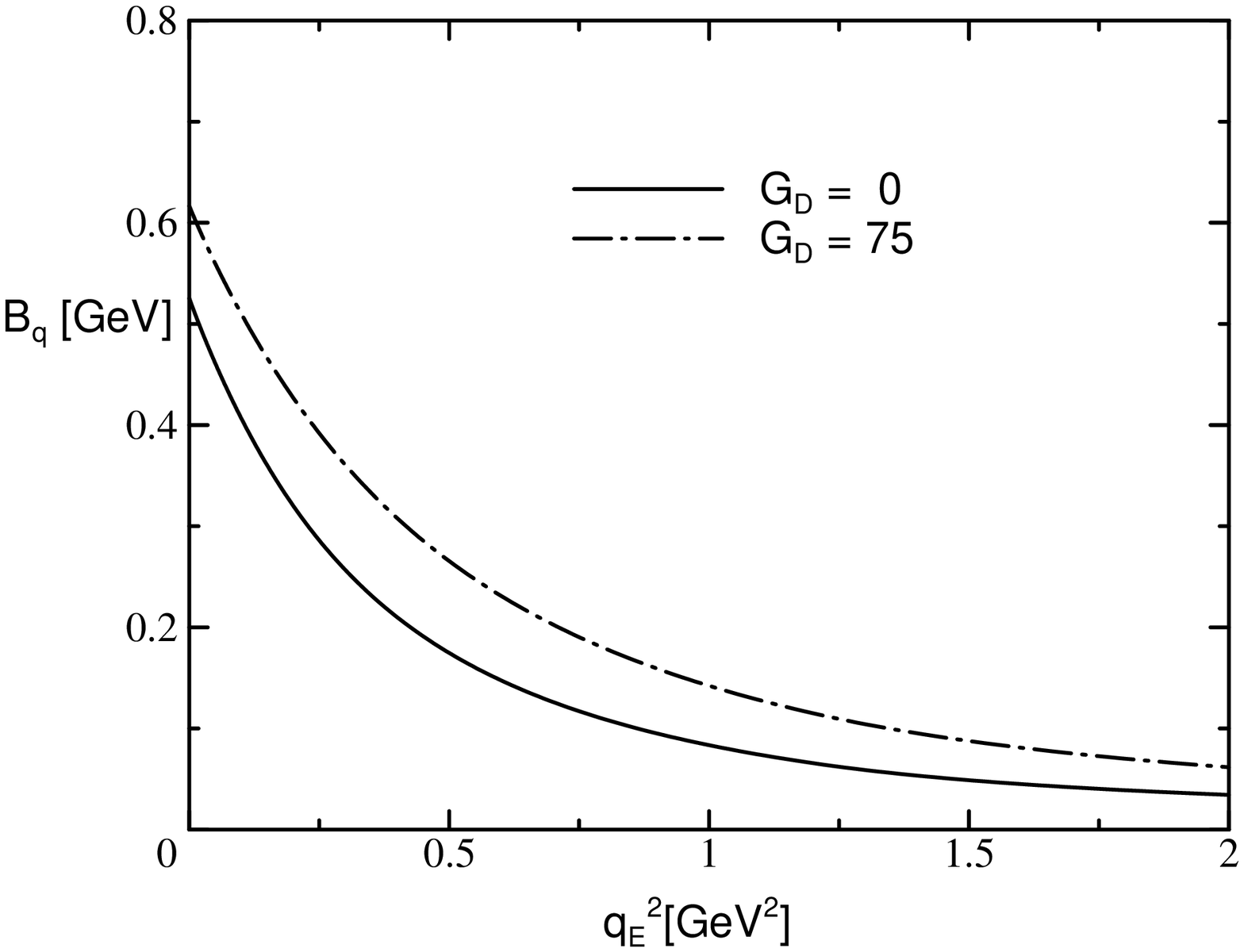}
&
\phantom{xxxxxxxxxxxxx}
&
\includegraphics[height=5cm]{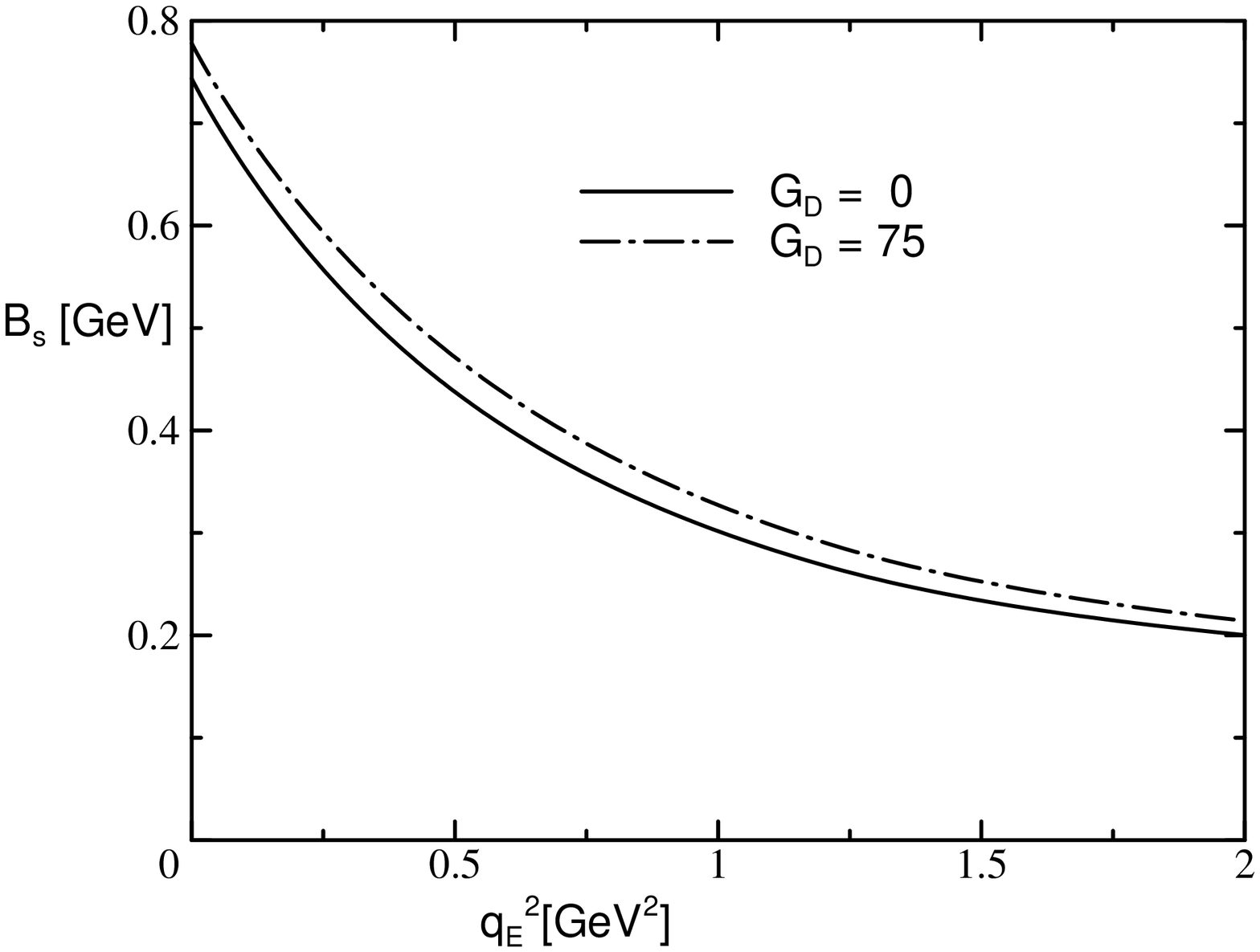}
\end{tabular}
\caption{\label{FIG:sd_q} $q_E^2$ dependencies of the mass function of the 
light (q) quark,  $B_q$, and the stranger (s) quark, $B_s$, 
for $G_D = 0$ and $75[\mbox{GeV}^{-5}]$ }
\end{figure*}

Generally the quark propagator is parametrized as 
\begin{eqnarray}
S_F(q) = \frac{i}{q \!\!\! /  A(q^2)-B(q^2)}. 
\end{eqnarray}
In the Landau gauge, it can be shown that the solution satisfies 
$A(-q_E^2)=1$. 
Then the SD equation becomes the integral equation only of the
mass function $B(q^2)$. 
Our numerical results of the mass function are  
shown in Fig.{\ref{FIG:sd_q}}. 
\par
The quark masses are renormalized as 
\begin{eqnarray}
m_q = Z_{m_q}^{-1}{m_q}_R  , \quad 
m_s = Z_{m_s}^{-1}{m_s}_R , 
\end{eqnarray}
where we take the renormalization condition as 
\begin{eqnarray}
&&\frac{\partial B_q(\mu^2)}{\partial {m_q}_R}\Bigg|_{m_{qR}=0}=1\\
&&\frac{\partial B_s(\mu^2)}{\partial {m_s}_R}\Bigg|_{m_{sR}=0}=1.
\end{eqnarray}
We define the  quark condensate as follows 
\begin{eqnarray}
  &&\langle \bar{\psi}(0) \psi(0) \rangle = 
  -\int _p \mbox{tr}[S_F^{R}(p)] 
  + \int_p \mbox{tr}[S_F^{\rm pert}(p)]
\end{eqnarray}
where
\begin{eqnarray}
&&S_F^{R}(p) = \frac{i}{p \!\!\! /    - B(p^2)}
\\
&&S_F^{\rm pert}(p) 
 = \frac{\partial S_F^R (p)}{\partial m_R} \Bigg{|} _{m_R = 0} m_R
\end{eqnarray}
\par

%%%%%%%%%%%%%%%%%%%%%%%%%%%%%%%%%%%%%%%%%%%%%%%%%%%%%%%%%%%%%%%%%%%
\subsection{\label{sec:BS}Bethe-Salpeter eqation}
%%%%%%%%%%%%%%%%%%%%%%%%%%%%%%%%%%%%%%%%%%%%%%%%%%%%%%%%%%%%%%%%%%%%
%
\begin{figure*}
\includegraphics[height=2cm]{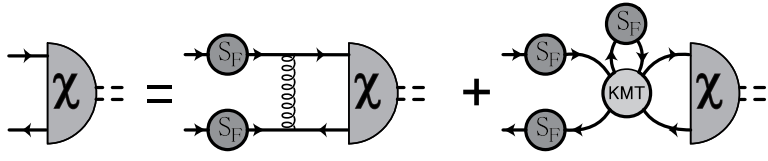}% Here is how to import EPS art
\caption{\label{fig:BSeq} The Bethe-Salpeter (BS) equation}
\end{figure*}
To treat the scalar and pseudoscalar mesons as 
quark-antiquark 
bound states, we use the homogeneous Bethe-Salpeter equation. 
The BS equation in momentum space becomes 
\begin{eqnarray}
%&&
S_F^{-1}(q_+)\chi(q;P)S_F^{ -1}(q_-)
%\nonumber\\
&=&-iC_F 
\int_k 
\bar{g}^2(-q^2,-k^2) 
i D^{\mu \nu}(q-k)\gamma_{\mu}\chi^R(k;P)\gamma_{\nu}\nonumber\\
&& - 2iG_D
\int_{p}\int_k 
w \left( -p^2 - q^2 - k^2 -\frac{P_B^2}{2}  \right)
\mbox{tr}^{\rm (DC)}[{S_F^R}(p)]
\nonumber \\
&&\times 
%\nonumber \\
%&&\times
\left\{
  \gamma_5\mbox{tr}^{\rm (DC)}[\gamma_5 \chi^R(k;P_B)] 
  + \mbox{tr}^{\rm (DC)}[\chi^R(k;P_B)]
\right\}
%\nonumber \\
\end{eqnarray}
where $q_{\pm} = q \pm \frac{P_B}{2}$, 
$\chi(k;P_B)$ denotes the BS amplitude and $P_B$ is the momentum of the meson. 
Like the SD equation, the $G_D$ term is antisymmetrized in the flavor space.
All the momentum integrals are regularized by the cut-off $\Lambda_{\rm UV}$, 
which  is chosen as $\Lambda_{\rm UV} = 100 [\mbox{GeV}]$ so that 
solutions of the BS equation do not depend on it. 
This equation is shown diagrammatically in Fig. \ref{fig:BSeq}.
\par
For the scalar meson, the BS amplitude can be parametrized in terms of 
four Dirac structures with appropriate flavor structures, 
\begin{eqnarray}
\chi^R(k;P)
&=&  \mbox{{\boldmath 1}}_{C} \frac{\lambda^{a}}{2}
\left( \phi_S^{a}(k;P) + \phi_P^{a}(k;P) k \!\!\! / 
+ \phi_Q^{a}(k;P) P \!\!\! /
\phantom{\frac{1}{2}}
%\right.
%\nonumber \\
%&&
%\left.
+\phi^{a}_T(k;P) \frac{1}{2}
(P \!\!\! / k \!\!\! / - k \!\!\!/  P \!\!\!/)
\right). 
\end{eqnarray}
%
%\par
%
The normalization condition of the BS amplitude is given by 
\begin{widetext}
\begin{equation}
i \int _{-q_{+}^2 \le \Lambda_{\rm UV}^2, -q_{-}^2 \le \Lambda_{\rm UV} ^2}
\frac{d^4 q}{(2\pi)^2}
\chi_{n_1m_1}(q;P)\bar{\chi}_{m_2n_2}(q;P)
P^{\mu}\frac{\partial}{\partial P^{\mu}}
\left\{{S_F}_{n_2n_1}^{-1}(q_+){S_F}_{m_1m_2}^{-1}(q_-) \right\}
=-2P^2,
\end{equation}
\end{widetext}
where the indices $m_1, n_1, \dots$ are combined indices in the 
color, flavor and Dirac spaces.
\par
In the numerical computation, we solve the BS equation in the 
Euclidean momentum region.
Then the physical solution, which corresponds to negative $P_E^2$, 
is obtained by extrapolation from the $P_E^2 > 0$ region. 
It can be done in the following way. 
First, we rewrite the Euclidean BS equation in the form 
\begin{eqnarray}
\label{directBS}
\phi_A(q;P_E) = \int _{k_E} M_{AB}(q_E;k_E;P_E) \phi_B(k;P_E)
\end{eqnarray}
where $\phi_A$ and $\phi_B$ denotes a set of amplitude. 
This equation should not have a solution at $P_E^2 > 0$ 
because, if there is one,  it is a tachyon solution. 
Therefore we instead solve an eigenvalue equation 
\begin{eqnarray}
\lambda(P_E^2) \phi_A(q_E;P_E) = \int _{k_E}M_{AB}(q_E;k_E;P_E) \phi_B(k_E;P_E)
\label{lambdaBS}
\end{eqnarray}
for a fixed $P_E^2 > 0$. 
Then we extrapolate the eigenvalue $\lambda$ to $P_E^2 < 0$ 
as a function of $P^2_E$ 
and search for the on-shell point $\lambda(-M_B^2) = 1$. 
We employ the quadratic function of $P_E^2$ 
for the extrapolation. 
%
%%%%%%%%%%%%%%%%%%%%%%%%%%%%%%%%%%%%%%%%%%%%%%%%%%%%%%%%%%%%%%%%%%%
\section{\label{sec:result}Numerical results}
%%%%%%%%%%%%%%%%%%%%%%%%%%%%%%%%%%%%%%%%%%%%%%%%%%%%%%%%%%%%%%%%%%%
%
%%%%%%%%%%%%%%%%%%%%%%%%%%%%%%%%%%%%%%%%%%%%%%%%%%%%%%%%%%%%%%%%%%%
\subsection{\label{sec:parameters}Parameters}
%%%%%%%%%%%%%%%%%%%%%%%%%%%%%%%%%%%%%%%%%%%%%%%%%%%%%%%%%%%%%%%%%%%
%
\begin{table}
\caption{\label{tab:mass} The numerical results of the BS equation. }
\begin{ruledtabular}
\begin{tabular}{ccc}
  & mass (MeV) &  \\ 
\hline
$\pi$   & $136$ & $f_{\pi}=95 [\mbox{MeV}]$   \\
$\eta$  & $515$ & mixing angle $-20.0^\circ$ \\
$\eta'$  & $982$ & mixing angle $-26.2^\circ$ \\
$K$  & $517$ & \\
\hline
$\sigma$  & $667$ & mixing angle $-68^\circ$ \\
%$\sigma$  & $685$ & mixing angle $-68^\circ$ \\
$a_0$  & $942$ & \\
%$a_0$  & $1045$ & \\
$f_0$  & $1336$ & mixing angle $-83.9^\circ$ \\
%$f_0$  & $1423$ & mixing angle $-83.9^\circ$ \\
$K_0^*$  & $1115 $ & \\
\end{tabular}
\end{ruledtabular}
\end{table}
\begin{table*}
 \caption{\label{tab:parameters}The values of the parameters 
  and the obtained observable quantities.  
 }
\begin{ruledtabular}
\begin{tabular}{cccccccc}
${m_q}_R(2[\mbox{GeV}]) [\mbox{MeV}]$ &
${m_s}_R(2[\mbox{GeV}]) [\mbox{MeV}]$ &
$\Lambda_{\mathrm{QCD}} [\mbox{MeV}]$ &
$t_{\mathrm{IF}}$ &
$t_0$ &
$G_D [\mbox{GeV}^{-5}]$ &
$\kappa [\mbox{GeV}^{-2}]$ &
\\
\hline
$4.5 $ &
$150 $ &
$600 $&
$ 0.204$&
$-6.89$&
$75 $&
$0.7 $
\end{tabular}
\end{ruledtabular}  
\end{table*}    
\par
In the present approach, there are seven input parameters:
the bare quark mass 
${m_q}_0$ for the up and down quarks,
in which we assume isospin symmetry $m_u = m_d$, 
 the current quark mass 
${m_s}_0$ for the strange quark,
 the scale parameter of QCD 
running coupling constant  $\Lambda_{\mathrm{QCD}}$, the 
infrared cutoff $t_{\mathrm{IF}}$, the smoothness parameter $t_0$, 
the strength parameter of the KMT $G_D$ and the parameter 
of the weight function of the KMT $\kappa$. 
We choose these parameters to reproduce the 
observables of the pseudoscalar mesons 
and then we apply them to the scalar mesons. 
\par
The quark masses ${m_q}_0, {m_s}_0$ are chosen 
so that the renormalized masses at the momentum scale $\mu = 2 [\mbox{GeV}]$ 
become the ${m_q}_R = 4.5[\mbox{MeV}]$ and ${m_s}_R = 150[\mbox{MeV}]$, 
respectively. 
The $\kappa$ parameter is taken as $\kappa = 0.7 [\mbox{GeV}^{-2}]$, 
which corresponds to the instanton of the average size $1/3[{\rm fm}]$. 
The other parameters $\Lambda_{\mathrm{QCD}}$, $t_{\mathrm{IF}}$, $t_0$, $G_D$
are chosen as the pseudoscalar meson masses $M_{\pi}$, $M_{\eta}$, 
$M_{\eta'}$ and the pion decay constant $f_{\pi}$ 
are fitted to the experimental values. 
The parameters that we use are 
$\Lambda_{\mathrm{QCD}} = 600[\mbox{MeV}]$, $t_0 = -6.89$, 
$t_{\mathrm{IF}} = 0.204$ and $G_D = 75 [\mbox{GeV}^{-5}]$. 
%
%\par
%
These parameters give 
$M_{\pi} = 136[\mbox{MeV}]$, $M_{\eta} = 515[\mbox{MeV}]$, 
$M_{\eta'} = 982 [\mbox{MeV}]$ and $f_{\pi} = 95 [\mbox{MeV}]$. 
These are in agreement with experimental values in less 
than $6\% $ of deviation.
Table \ref{tab:parameters} summarizes all the values of the parameters. 
\par
Although $\Lambda_{\mathrm{QCD}}$ is somewhat larger than the standard value 
$\Lambda_{\mathrm{QCD}} = 100 \sim 300 [\mbox{MeV}] $, 
a large $\Lambda_{\mathrm{QCD}}$ is necessary in the ILA to generate 
the desired  dynamical chiral symmetry breaking (D$\chi$SB). 
%
%\par
%
It is known that by using the running coupling constant obtained by
the higher loop calculation, the smaller $\Lambda_{\rm QCD}$ 
gives rise to the correct size of the DCSB. 
On the other hand, the Wilsonian non-perturbative renormalization 
group analysis has shown that 
the effects beyond the ladder approximation also 
reduce $\Lambda_{\rm QCD}$ \cite{AokiNPRG}. 
%
%%%%%%%%%%%%%%%%%%%%%%%%%%%%%%%%%%%%%%%%%%%%%%%%%%%%%%%%%%%%%%%%%%%
\subsection{\label{sec:SDresult}Solution of the SD equation}
%%%%%%%%%%%%%%%%%%%%%%%%%%%%%%%%%%%%%%%%%%%%%%%%%%%%%%%%%%%%%%%%%%%
%
The numerical solutions of the SD equation are shown in Fig.{\ref{FIG:sd_q}}. 
The values of the quark condensates 
and mass function at $P_E^2 = 0$ are 
given  in the Table \ref{TAB:sdResult}. 
\begin{table}
 \caption{\label{TAB:sdResult}  
  The values of the mass function at $P_E^2 = 0$ 
and the quark condensates. 
 }
\begin{ruledtabular}
\begin{tabular}{ccc}
$B_q(0)$ &
$B_s(0)$ &
$-\langle \bar{q}q \rangle_R^{1/3} $
\\
\hline
$0.616 [\mbox{GeV}]$ &
$0.778 [\mbox{GeV}]$ &
$0.259 [\mbox{GeV}]$
\\
\end{tabular}
\end{ruledtabular}
\end{table}
%
%
%%%%%%%%%%%%%%%%%%%%%%%%%%%%%%%%%%%%%%%%%%%%%%%%%%%%%%%%%%%%%%%%%%%
\subsection{\label{sec:resultScalar}Solution of the BS equation 
for the Scalar mesons}
%%%%%%%%%%%%%%%%%%%%%%%%%%%%%%%%%%%%%%%%%%%%%%%%%%%%%%%%%%%%%%%%%%%
%
As mentioned above, since 
the parameters of this approach have been chosen 
using the observables of the pseudoscalar mesons, 
the following numerical results of the scalar mesons 
are parameter free predictions.
\begin{figure}
\includegraphics[height=6cm]{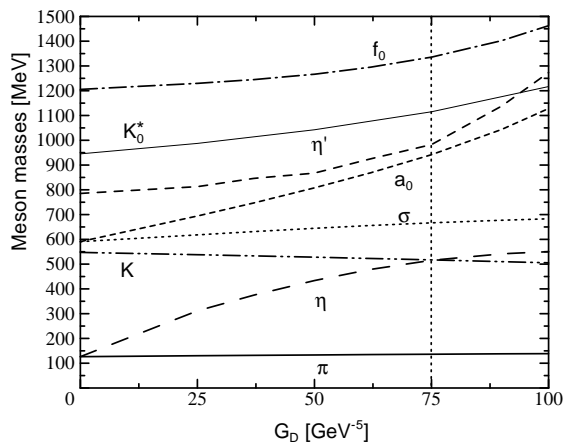}
\caption{\label{FIG:M_nonet} Dependence of the mass spectrum of the scalar
 and pseudoscalar meson nonet on the strength of the KMT interaction. }
\end{figure}
\par
Let us now start the discussion of the solutions of the BS equation. 
Our numerical results for the scalar mesons are 
summarized in Table \ref{tab:mass}. The dependence of the 
mass spectra  on the strength of the KMT interaction 
is shown in the Fig.{\ref{FIG:M_nonet}}. 
\par
First, the dependencies of the masses of $\sigma$ and $a_0$ 
on $G_D$ look qualitatively same as the NJL results shown in \cite{NOTU2003}. 
In the NJL calculation, the parameters are chosen so as to 
reproduce the $M_{\pi}$ and $f_{\pi}$ at each $G_D$. 
In contrast, in the ILA approach we change $G_D$ independently. 
However, since $M_{\pi}$ and $f_{\pi}$ depend weakly on 
$G_D$, the results of ILA show similar behavior as those of NJL. 
We note that the mass of $a_0$ grows as $G_D$ increases, while 
the $\sigma$ mass is almost constant. 
\par
The $\sigma$ meson mass is predicted as about $670 [\mbox{MeV}]$. 
This rather small $\sigma$ meson mass is interesting. 
In the case of the NJL model, 
the $\sigma$ meson mass is determined to be close to 
twice of the dynamical quark mass. 
On the other hand, in the ILA approach 
the value of the mass function at $q_E^2 = 0 $, $B_q(q_E^2=0)$ 
is about $616 [\mbox{MeV}]$, 
which is comparable to the $\sigma$ meson mass. 
Recently, Maris calculated the $\sigma$ meson mass in the
SD and BS approach with the infrared enhanced gluon propagator 
\cite{Maris2002} and the obtained mass is $671 [\mbox{MeV}]$.
Although there are such differences, the properties of 
the physical observables agree in these calculations. 
\par
Concerning the $a_0$ meson,
we obtain $M_{a_0} = 942 [\mbox{MeV}]$. This result is comparable 
to the experimental value $984.8 \pm 1.4 [\mbox{MeV}]$. 
We obtain significant mass splitting between the $\sigma$ and $a_0$,
 about $275[\mbox{MeV}]$. 
We conclude that the observed $\sigma-a_0$ mass splitting can be 
explained as the $U_A(1)$ symmetry breaking effects.
Recent study of the light flavor scalar mesons in the instanton 
liquid model showed similar result \cite{Schafer2003}.
\par
The obtained mass of $f_0$ is $1336$ MeV. We therefore identify
this state with $f_0(1370)$ whose T-matrix pole position is 
$(1200 - 1500) - i (150 - 250)$ MeV.
The calculated mass of $f_0$ is about $400$ MeV larger than that of
$a_0$. The mixing angle of $f_0$ is $-83.9^\circ$ and it means that
this state is close to the flavor octet state. In such a case,
the effects of the KMT term on $f_0$ state is almost same as that on 
$a_0$ and therefore the origin of the mass difference between 
$f_0$ and $a_0$ is considered to be mostly the symmetry breaking 
effects by the strange quark mass.
\par
The $f_0(980)$ state is observed between $\sigma(600)$ and $f_0(1370)$.
From the present study,
it seems unlikely that $f_0(980)$ is a simple $q \bar q$-bound state.
Literatures have suggested that $f_0(980)$ may consist of $q^2 \bar q^2$
\cite{Jaffe1987,AJ2000,BFSS1999,CT2002} or may be a $K \bar K$ molecular state
\cite{WI1990}. Those studies often indicate that the $a_0(980)$
is also a four-quark state. It conflicts with our picture.
Further study is necessary to make the situation clearer. 
\par
The dependences of the scalar meson spectrum on the strength of 
\UA\ breaking interaction are understood as follows. 
When $G_D = 0$,  $\sigma$ meson is ideal mixing state 
$\frac{1}{\sqrt{2}}(u\bar{u} + d\bar{d})$. As the $G_D$ becomes large, 
the $\sigma$ approaches the flavor singlet state 
$\frac{1}{\sqrt{3}}(u\bar{u} + d\bar{d} +s\bar{s})$. 
The $\sigma$ meson mass seems to become large 
because of the increase of the strange component. 
On the other hand, the KMT interaction is 
attractive for the flavor singlet state, 
consequently the $\sigma$ meson mass hardly change. 
In the case of $a_0$, 
since the KMT interaction 
dose not induce the flavor mixing,  $a_0$ is fixed on 
the $\frac{1}{\sqrt{2}}(u\bar{u} + d\bar{d})$ state. 
As $G_D$ becomes large, the  $a_0$ mass increase 
by the repulsive KMT interaction. 
As $G_D$ becomes large, the  $f_0$ changes 
from $s\bar{s}$ to 
$\frac{2}{\sqrt{3}}(u\bar{u} + d\bar{d} -2s\bar{s})$. 
The decreasing of the strange component 
and the repulsive KMT interaction compete.  
At the small $G_D$, they are balanced and 
the $f_0$ mass increases at the large $G_D$ where 
the flavor of the $f_0$ is sufficiently mixed. 
\par
\begin{figure}
\includegraphics[height=6cm]{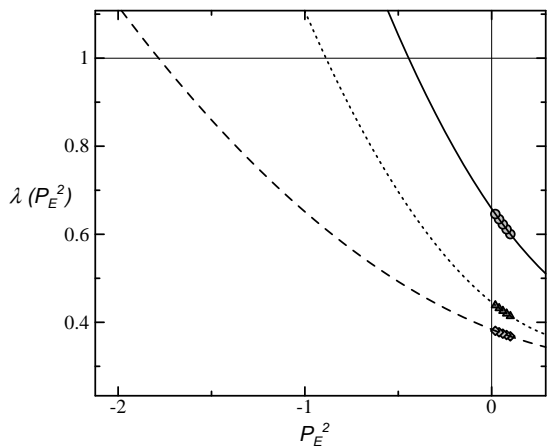}
\caption{\label{FIG:extra} Extrapolation of the eigenvalue $\lambda$ 
with respect to $P_E^2$ 
for $\sigma$,  $a_0$ and $f_0$ mesons. 
The solid line shows the extrapolation of the $\sigma$ and 
the doted line shows the extrapolation of the $a_0$ and the dashed line 
shows the extrapolation of the $f_0$.}
\end{figure}

There is a shortcoming in the present approach.
The solution of the BS equation is 
obtained at the Euclidian momentum region. 
To obtain the physical mass and mixing angle, 
we have to extrapolate the solution from 
the space-like to the time-like region. 
This is carried out by extrapolating the eigenvalue function 
$\lambda(P_E^2)$ in Eq. (\ref{lambdaBS}) to negative $P_E^2$ until 
it hits $\lambda(P_E^2)=1$. 
The graphs of the extrapolation about the masses of 
the $a_0$, $\sigma$, $f_{0}$ 
are shown in the Fig.$\ref{FIG:extra}$
We therefore anticipate moderate ambiguity in the extracted
masses especially when the mass is large.

\begin{figure}
\includegraphics[height=6cm]{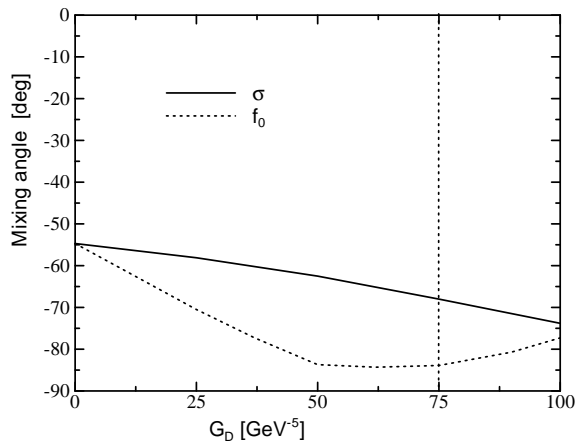}
\caption{\label{FIG:s_mixing} Depnendence of the mixing angle of the $\sigma$
 and $f_0$ meson on the strength of the KMT interaction.  }
\end{figure}

Next, we consider the mixing angles. 
We introduce the matrix elements $S^8$ and $S^0$ which are defined by 
\begin{eqnarray}
S^a &=& \int d^4x \langle 0 | \bar{q} \frac{\lambda^a}{2} q (x)
| \mbox{scalar meson state} \rangle \\
&=& \mbox{tr} \left[\bar{\chi}^R(0;P)\frac{\lambda^{a}}{2} \right]
\end{eqnarray}
Since these S values extract the particular flavor 
component of $\phi_S$ which is 
the main component of the BS amplitude at the origin, 
we employ $S^8$ and $S^0$ to determine the 
ratio of the octet and 
the singlet components. 
Accordingly we define the mixing angles 
of the scalar mesons as 
\begin{eqnarray}
&&\tan \theta_{\sigma} = - \frac{S_0^{\sigma}}{S_8^{\sigma}} \\
&&\tan \theta_{f_0} = \frac{S_8^{f_0}}{S_0^{f_0}}. 
\end{eqnarray}
The results are summarized 
in Table \ref{tab:mass} and 
Fig. \ref{FIG:s_mixing}. 

As $G_D$ increases, the mixing angle moves towards $-90^{\circ}$, 
where $\sigma$ becomes the purely flavor singlet state: 
$\frac{1}{\sqrt{3}}(u\bar{u} + d\bar{d} + s\bar{s})$.
The obtained angle at $G_D=75[\mbox{GeV}^{-5}]$ is $-68.0^\circ$ and is 
slightly smaller than the result of the 
NJL model, $-77.3^\circ$. This angle corresponds to 
about $5\%$ mixing of the strangeness  component 
in $\sigma$.

%%%%%%%%%%%%%%%%%%%%%%%%%%%%%%%%%%%%%%%%%%%%%%%%%%%%%%%%%%%%%%%%%%%
\subsection{\label{sec:resultStrange}Solution of the BS equation for
the Strange meson}
%%%%%%%%%%%%%%%%%%%%%%%%%%%%%%%%%%%%%%%%%%%%%%%%%%%%%%%%%%%%%%%%%%%

In this section, we discuss the strange scalar meson $K_0^*$. 
Here we employ an approximation in order to 
avoid the technical difficulty
coming from the ambiguity in defining the center of mass coordinate as 
this meson 
consists of the a strange quark and a nonstrange quark. 
Instead of treating the asymmetric  BS equation, 
we solve the symmetric BS equation for 
the quarks of mass ${m_q+m_s\over 2} = 77.25 [\mbox{MeV}]$.
We apply this approximation to the kaon and obtained the 
reasonable kaon mass, $m_K = 494 - 498$ MeV.
\par
The results are summarized in Table \ref{tab:mass} 
 and Fig. \ref{FIG:M_nonet}. 
The obtained $K_0^*$ mass is $1115$ MeV, which is about $173$ MeV 
larger than that of $a_0$ and about $221$ MeV smaller than that of $f_0$.
If all the $a_0$, $K_0^*$ and $f_0$ states are assumed to be the flavor
octet $q \bar q$ states, the simple quark model calculation predicts the
Gell-Mann-Okubo mass formula. Our results deviate from both the linear
mass formula, $3 (M_{f_0} - M_{K_0^*}) = M_{K_0^*} - M_{a_0}$ and 
the quadratic mass formula, 
$3 (M_{f_0}^2 - M_{K_0^*}^2) = M_{K_0^*}^2 - M_{a_0}^2$.
\par
Unfortunately, the corresponding  light $I = 1/2$ scalar meson
is not observed. The observed mass of the $K_0^*(1430)$ is 
$1412 \pm 6$ MeV and larger than our result of $M_{f_0}$.
It therefore seems to be rather difficult to identify $K_0^*(1430)$
with our $K_0^*$ state.
We do not consider $K_0^*(1430)$ as a member of the light scalar nonet states.
We hope the re-analysis of the experimental data using the chiral effective 
model including the effects of the \UA\-breaking interaction will shed light 
on the problem of the missing state in the $I = 1/2$ scalar channel.

%%%%%%%%%%%%%%%%%%%%%%%%%%%%%%%%%%%%%%%%%%%%%%%%%%%%%%%%%%%%%%%%%%%
\section{\label{sec:conclusion}Summary and Conclusions}
%%%%%%%%%%%%%%%%%%%%%%%%%%%%%%%%%%%%%%%%%%%%%%%%%%%%%%%%%%%%%%%%%%%
\par
We have studied the spectrum of the
 light scalar nonet mesons using the improved ladder approximation (ILA) 
of QCD with the Kobayashi-Maskawa-'t~Hooft (KMT) \UA\ breaking interaction. 
We choose parameters to reproduce the masses and decay constants of the 
pseudoscalar nonet mesons and apply those to the scalar nonet mesons. 
\par
The ILA of QCD is  an approximation that is consistent with chiral 
symmetry and consists of the rainbow approximation of the 
Schwinger-Dyson equation and the ladder approximation 
of the Bethe-Salpeter equation. 
Using this approach, we analyze the scalar meson spectrum 
quantitatively. 
\par
We have obtained $M_{\sigma} = 667$ MeV, $M_{a_0} = 942$ MeV and 
$M_{f_0} = 1336$ MeV. 
Considering the ambiguity due to the extrapolation from the Euclid 
momentum, they are in good agreement with the observed masses
of $\sigma(600)$, $a_0(980)$ and $f_0(1370)$, respectively.
We therefore consider that these states are the members of the 
light scalar meson nonet.
This identification is different from the conventional one. 
The key ingredient 
is the instanton-induced \UA\ breaking interaction.
It gives rise to the symmetry breaking and flavor mixing 
effects on the scalar mesons 
as well as the pseudoscalar mesons.
\par
We have obtained the strangeness content in 
the $\sigma$ meson of about $5\%$. 
This $s\bar{s}$ mixing may be tested, for instance, by analyzing the 
$\pi$ $\pi$ decays of heavy mesons carefully. 
\par
The obtained $K_0^*$ mass is $1115$ MeV. 
The corresponding state is not observed. 
The observed $K_0^*(1430)$ is heavier than our result of $M_{f_0}$ and 
therefore we do not include $K_0^*(1430)$ in the light scalar nonet.
\par
It should be noted that the scalar isoscalar state can be mixed 
with the scalar glueball state.  Such effect is not taken 
into account here. There is another important effect to be discussed.
The light scalar nonet can be generally coupled to the intermediate 
states composed of two pseudoscalar mesons rather strongly.
If the light $q^2 \bar q^2$ states exist, this coupling effects 
should be very important.
The extension of the Bethe-Salpeter equation to the $q^2 \bar q^2$ 
system seems to be rather difficult. One way is to apply the 
bilocal bosonisation technique \cite{RCSI1994}.
Another direction of the further study is the extension to the 
finite temperature and/or density \cite{MRST2001}. 
In the present study, the mass of the $I = 1/2$ scalar meson is not 
reproduced well. In order to make the structure of the light scalar mesons
clearer, the analysis of the experimental data using the framework 
including the \UA\ breaking interaction is necessary. 
%
%
%%%%%%%%%%%%%%%%%%%%%%%%%%%%%%%%%%%%%%%%%%%%%%%%%%%%%%%%%%%%%%%%%%%%%%
%                                                                    %
% Acknowledgment                                                     %
%                                                                    %
%%%%%%%%%%%%%%%%%%%%%%%%%%%%%%%%%%%%%%%%%%%%%%%%%%%%%%%%%%%%%%%%%%%%%%
%
%
\section*{Acknowledgment}
This work is supported in part by the Grant-in-Aid for Scientific Research
(B) 15340072 of the Ministry of Education, Science, Sports and 
Culture of Japan, and by the Cooperation Research Program
of the Showa Pharmaceutical University.
%
%
%%%%%%%%%%%%%%%%%%%%%%%%%%%%%%%%%%%%%%%%%%%%%%%%%%%%%%%%%%%%%%%%%%%%%%
% reffereces 
%%%%%%%%%%%%%%%%%%%%%%%%%%%%%%%%%%%%%%%%%%%%%%%%%%%%%%%%%%%%%%%%%%%%%%

\end{document}